\documentclass[12pt,a4paper,twocolumn,singlespacing,fleqn,usenatbib]{mnras}
\usepackage{ae,aecompl}
\usepackage{amsmath}
\usepackage{natbib}
\usepackage{graphicx}
\usepackage[utf8]{inputenc}

\title{Evolution of Hubble parameter from Pantheon+  data and comparison of cosmological models using cosmic chronometers}
\author[Ardra Edathandel Sasi]{Ardra Edathandel Sasi$^{1}$
and Moncy Vilavinal John$^{1}$\thanks{moncyjohn@yahoo.co.uk}\\
$^{1}$School of Pure and Applied Physics, Mahatma Gandhi University, Kottayam, India}
\pubyear{\the\year{}}

\date{\today}

\begin{document}
\label{firstpage}
\pagerange{\pageref{firstpage}--\pageref{lastpage}}
\maketitle

\begin{abstract}
The evolution of the Hubble parameter $H(z)$ with redshift $z$ is estimated from the Pantheon+ data of Type Ia supernovae, for the $\Lambda$CDM model and the three special cases of the eternal coasting (EC) cosmological model with three different spatial geometries. The scatter associated with $H(z)$ is seen to grow markedly with redshift. This behaviour, which is deduced directly from the SNe Hubble diagram, raises the question of whether the universe is undergoing a stochastic expansion, which scenario can offer an explanation for the Hubble tension in cosmology. From the estimated $H(z)$ values, the present value of the Hubble parameter $H_0$ is evaluated  for each of these models through regression, and the scatter using the Monte Carlo method. Bayesian  comparison between these models is carried out using the data of 35 cosmic chronometers (CC). The comparative study favours the $\Lambda$CDM model, with some strong evidence. However, exclusion of four outlier CC data points with small errorbars leads to large reduction in the Bayes factor value. The unusually large value of Bayes factor obtained while using the full set of CC data raises some concerns about its tension with other data, such as that of the SNe Ia. While using the remaining 31 CC data points, it is observed that the resulting Bayes factor still favours the $\Lambda$CDM model, but with a much smaller value of the Bayes factor.  When EC models are compared among themselves, the $\Omega = 2$ model has strong evidence than the $\Omega = 1$ (also known as $R_h = ct$) and the $\Omega = 0$ (Milne-type) models.

\end{abstract}


%%%%%%%%%%%%%%%%%%%%%%%%%%%%%%%%%%%%%%%%%%%%%%%%%%%%%%%%%%%

\section{Introduction}
Though  cosmology as a modern science has its history lasting only for roughly one century, it could provide us with a fairly accurate picture of the evolution of our universe, from  its hot very early phase to the present state, that took place during  the past 13.8 billion years or more. An almost complete and satisfactory theoretical account of this evolution is given by the $\Lambda$CDM model \citep{peebles1984tests,peebles1993principles}. This model conceives the universe as  containing ordinary matter ($\sim 5$ per cent), a larger amount of unseen dark matter ($\sim 25$ per cent) and a much larger amount of still mysterious dark energy ($\sim 70$ per cent). A host of observational data, such as those related to the relative abundance of light elements in the universe (big bang nucleosynthesis), the temperature fluctuations in the cosmic  microwave background (CMB) radiation (the CMB power spectrum), formation of large scale structures such as clusters and superclusters of galaxies, voids, etc., univocally provide strong support to this model.  While the $\Lambda$CDM  model is largely successful on several such fronts, there lingers  even today some problems and tensions in this  model, such as the Hubble tension, $\sigma_8$ tension, coincidence problem, etc. \citep{di2022challenges}.

An alternative to this model, named the eternal coasting (EC) cosmological  model \citep{john1996modified,john1997low,john2000generalized,melia2012r} where the scale factor of the universe varies linearly with time ($a \propto t$), has gained increased attention in recent years. [For a review, see \citep{casado2020linear}.]  The most characteristic feature of the model in \citep{john1996modified,john1997low,john2000generalized} is that it has  all  components, such as the ordinary and dark matter, dark energy, etc., varying as $a^{-2}$.  To maintain this, there will be continuous creation of matter/dark matter at the expense of dark energy, leading to a constant ratio between  matter density and dark energy density. Hence it  is devoid of any  coincidence problem.  The earliest work in \citep{john1996modified,john1997low}, which is specifically a bouncing and coasting model, has closed spatial geometry ($k=+1$) and contains an additional negative energy density  varying as $a^{-4}$, though it disappears very early. But the model  in \citep{john2000generalized} is more general and considers all the three possible 3-geometries.  It was explicitly shown   that these  models will have none of the cosmological problems which plagued  the  FLRW models.  The evolution of temperature in this `early-dark energy' model was shown  \citep{john1997low} to be almost  the same as that in the standard model. A special case ($k=0$) of this model  is studied extensively under the title $`R_h=ct$ model', where it was shown  capable of explaining several  observational data related to the expansion  of the universe \citep{melia2012r}. However, in the $R_h=ct$ version, there is no definite  inventory of  matter/energy components for the cosmic fluid or there is no clear-cut prescription for the variation of density parameters of  such components \citep{john2019r}, as in \citep{john2000generalized}.

In 1998,  with the release of  the apparent magnitude-redshift ($m$-$z$) data of Type Ia Supernovae (SNe Ia) \citep{riess1998observational,perlmutter1999measurements}, it was found that the expansion of the present universe is in marked deviation from the decelerating Friedmann solutions that reigned till then.  SNe Ia are ideal standard candles with which the distances to the  universe up to redshifts $z\sim 2-3$ are reliably estimated and these are still considered to be the primary source of information in our understanding of the universe. The strong claim made by the $\Lambda$CDM model that the present  universe is accelerating was, however, not  undisputed. In  a first ever Bayesian comparison  of cosmological models \citep{john2002comparison},  the $\Lambda$CDM model was compared with the EC model using the $m$-$z$ data of SNe Ia and the angular size-redshift data of galaxies. The results showed that there is only some marginal advantage for the $\Lambda$CDM  model over the EC model in accounting for these data. Later, some model-independent analyses using SNe Ia data \citep{john2005cosmography,john2010bayesian} showed that there is significant probability for the deceleration parameter $q_0$ to be zero, thus providing credence to the linear coasting model. 

 Recently, in view of the looming Hubble tension in cosmology \citep{di2021realm}, several novel cosmological probes are developed  to investigate the history of cosmic evolution. (For a  review, see \citep{moresco2022unveiling}.) A prominent  observation that belongs to this category is that of the cosmic chronometers (CC) that help to evaluate  the Hubble parameter $H(z)$   in a  model-independent manner. Astrophysical objects that can serve as CC are passively evolving galaxies whose redshift determination can be done with extreme accuracy. Such objects  allow us to trace the differential age evolution of the universe across a wide range of cosmic times. Notable applications of cosmic chronometers are in the estimation of the present value of the Hubble parameter $H_0$, estimation of other cosmological parameters, comparison with other probes, comparison of different cosmological models, etc.

In this work, we first make use of the fact that in a particular cosmological model,  each SN Ia with redshift $z$ gives a  value of $H(z)$. If there is no error in the measurement of $m$ and $z$, we get a definite value of $H(z)$.  When there is a nonzero error, one can find a probability distribution for $H(z)$ at $z$. Though these results are model-dependent,  the values of expansion rates of the universe at various cosmic times, extracted from SNe Ia would be  valuable information in the context of the Hubble tension. We perform this computation for the $\Lambda$CDM and EC models and in both cases, the $H(z)\pm \sigma_{H}$ versus $z$ plots were made. We notice the growth in the scatter in $H(z)$ with $z$ while doing Gaussian progress regression (GPR). This again is relevant information, when Hubble tension is concerned. It raises the question whether there is some inherent fluctuation in the expansion rate of the universe \citep{berera1994stochastic,sivakumar2001stochastic,john2003classical}, which subsides with the passage of cosmic time. This can also be a potential explanation to the disparity between the predicted and local measurements of $H_0$, as in Hubble tension. In the second part of the work, we use Monte Carlo simulations to randomly sample noisy data and evaluate the uncertainty in the parameter $H_0$, after estimating it through regression.  In each of these models we get different values of $H_0\pm \sigma_{H_0}$. Such theoretical predictions of all these four models are then subjected to Bayesian model comparison using the data of 35 cosmic chronometers \citep{moresco2024addressing}, with appropriate prior probabilities obtained from the above evaluation of $H_0$. The significance of the resulting Bayes factors are discussed. We extend this study to see whether the same results  follow if we eliminate a few outlier data points with small errorbars. By eliminating four such points from the CC data,  a huge difference in the Bayes factors is observed, which we argue as indicating a possible inconsistency between the SNe Ia data and the CC data when the latter is taken in full. We also find that a  Bayesian comparison of these models, as performed  in \citep{melia2013cosmic,melia2018model}, has certain flaw in choosing the prior probabilities.

The plan of the paper is as follows: In the next section, we outline the evaluation of the Hubble parameter for the $\Lambda$CDM model and the EC models, using the SNe data and then the evaluation of $H_0$ in each of these cases, from the resulting $H(z)$ diagram. Here we also present our results for the scatter in the $H(z)$ diagram. Section 3 deals with the Bayesian model comparison of the four cosmological models, with the full and modified sets of CC data. The last section comprises our conclusion.

\section{$H(z)$ in the $\Lambda$CDM and the EC models from SNe Ia data}

The $\Lambda$CDM model has flat spatial geometry $(k = 0)$. The density parameter corresponding to  matter (including dark matter) is $\Omega_m$ and that corresponding to dark energy is $\Omega_{\Lambda}$, such that \(\Omega_m + \Omega_{\Lambda} = 1\). The model grants expression for luminosity distance as
       \begin{equation}
            D_L= \frac{c}{H(z)}(1+z)I(z), 
       \end{equation}
       where
       
       \begin{equation}
           I(z) = \int_0^z [(1+z)^2(1+\Omega_m z)-z(2+z)\Omega_{\Lambda}]^{\frac{1}{2}} \,\mathrm{d}z ,
       \end{equation}

       Using SN Ia data, one may estimate the value of the Hubble parameter $H(z_j)\equiv H_j$ for each of the supernova at $z_j$,  using the expression for $D_L$. When there is nonzero error in measurements, one can  evaluate $H_j\pm \sigma_{H_j}$ versus $z_j$, corresponding to each of the supernova  and plot the $H$ versus $z$ diagram. For this, we may use the Chi-squared statistic, which uses

\begin{equation}
\chi_j^2=   \frac{(m_{p,j} -m_{o,j})^2}{\sigma_j^2}. \label{eq:chi2prob}
\end{equation}
Here $m_{p,j}$ and $m_{o,j}$ are respectively the predicted and observed values of the apparent magnitude of supernova, each corresponding to $z_j$, with $H_j$ as one of the free parameters. The  $\chi_j^2$ computed for  the $j^{th}$ SNe Ia  can be used to obtain 

\begin{equation}
P(H_j|D,M_1)=\frac{\exp(-\chi_j^2/2)}{\int_{-\infty}^{+\infty}\exp(-\chi_j^2/2)\mathrm{d}H_j}. \label{eq:chi2prob1}
\end{equation}
 
\begin{figure}
%\begin{minipage}{0.7 \textwidth}
    \centering
    \includegraphics[width=1\linewidth]{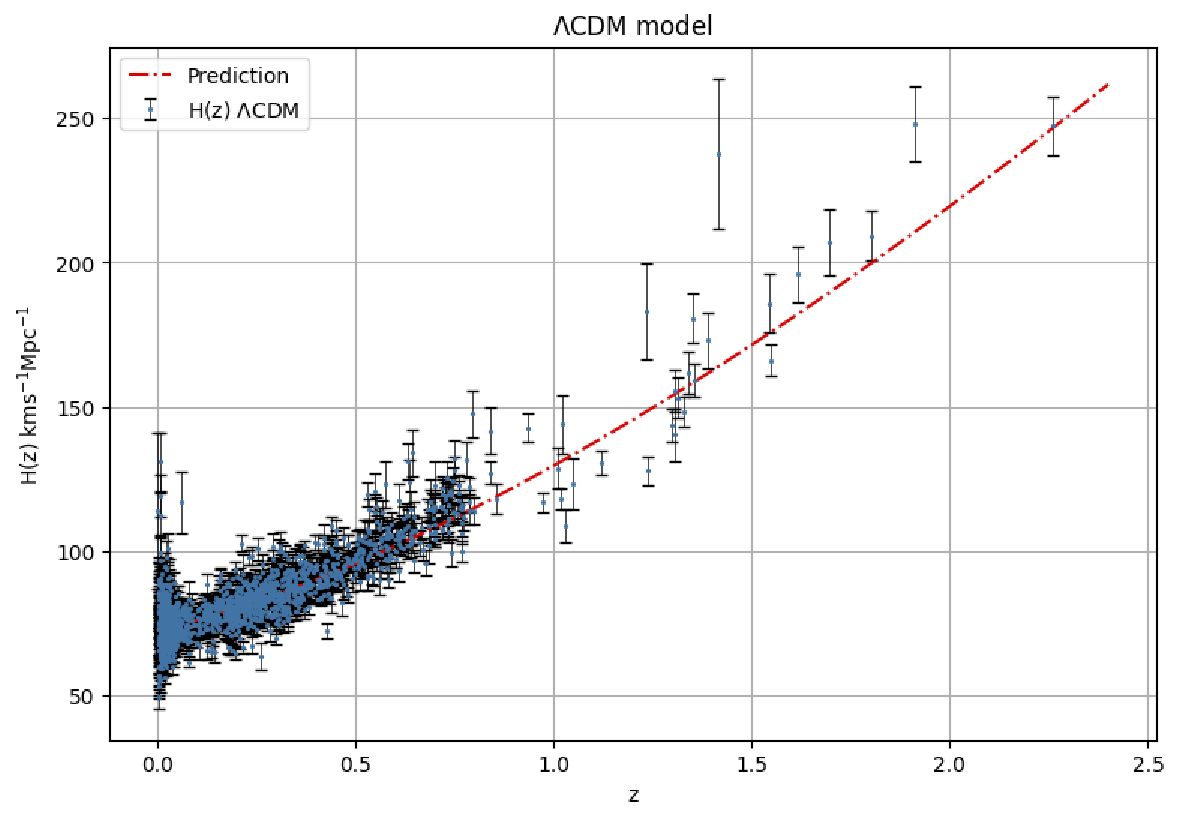}
    \caption{Hubble parameter evolution in the $\Lambda$CDM model, red line giving the theoretical prediction of the model as in equation (\ref{eq:Hzlcdm}). The regession and Monte Carlo methods give a value of Hubble constant $H_0 = 72.391 \pm 0.053$ km s$^{-1}$ Mpc$^{-1}$.}
    \label{fig:Figure2}
%\end{minipage}
\end{figure}

 This gives a probability distribution function for $H_j$, given the data $D$ and the validity of the model $M_1$, provided we give fixed values to the parameters other than $H_j$. From this pdf, its mean values $\bar{H}_j$ and the standard deviation $\sigma_{H_j}$ can be evaluated. These mean values  will depend on the free parameters of the model. In the $\Lambda$CDM model, since we have flat geometry, the model has a fixed value $\Omega =1$, but the matter density parameter $\Omega_m$, the Hubble paramater $H_j$ and the absolute magnitude $M$ of an SN Ia are free parameters in this model. In our calculations, we have fixed $\Omega_m=0.315\pm 0.007$  \citep{lahav2024cosmological} and $M=-19.3$ as  fiducial values and varied only $H_j$. A plot showing the mean and standard deviation of the Hubble parameter corresponding to each of the supernovae in the $\Lambda$CDM model  is given in Fig. \ref{fig:Figure2}.

In the $\Lambda$CDM model,  one has an expression for the variation of the Hubble parameter, given by

\begin{equation}
H(z)=H_0 \sqrt{\Omega_m(1+z)^3+\Omega_{\Lambda}} \label{eq:Hzlcdm}
\end{equation}
 We estimate the parameter $H_0$ in this expression  (again fixing $\Omega_m=0.315\pm 0.007$) by regression and evaluate the uncertainty in the parameter $H_0$ using Monte Carlo method. This is with the aid of the $H$-$z$ diagram we constructed out of the Pantheon+ data. The resulting value is $H_0=72.391 \pm 0.053$ km s$^{-1}$ Mpc$^{-1}$ and the curve thus obtained is overplotted in Fig. \ref{fig:Figure2}.

On the otherhand, for the the EC model,  the time evolution of scale factor  can be given for all the three space geometries ($k=0,\pm 1$) as

\begin{equation}
a=\alpha ct, \label{eq:mct}
\end{equation}
where

\begin{equation}
 \alpha =\sqrt{\frac{k}{\Omega -1}}.
\end{equation}
This $a(t)$ is the solution of the Friedmann equations when all energy densities vary as $a^{-2}$, which was the modified Chen-Wu ansatz proposed in \citep{john2000generalized}. (This in turn implies zero gravitational charge   $\rho c^2+3p=0$.) It may be noted that here one can take $\alpha =1$ for each of $\Omega =$ -1, 0 and +1. The luminosity distance in these cases is

\begin{equation}
D_L= \frac{\alpha c}{H(z)}(1+z)^2\sin n\left[ \frac{1}{\alpha} \ln (1+z) \right], \label{eq:D_L_local}
\end{equation}
where $\sin n(x) = \sin (x)$ for $k=+1$, $\sin n(x) = x$ for $k=0$ and $\sin n(x) = \sinh (x)$ for $k=-1$. Note that  the above procedure we adopted in the $\Lambda$CDM model can be used to obtain the mean values $H_j$ and the standard deviation $\sigma_{H_j}$ also in this model.

\begin{figure}
    \begin{minipage}{0.45\textwidth}
        \centering
 \includegraphics[width=1\linewidth]{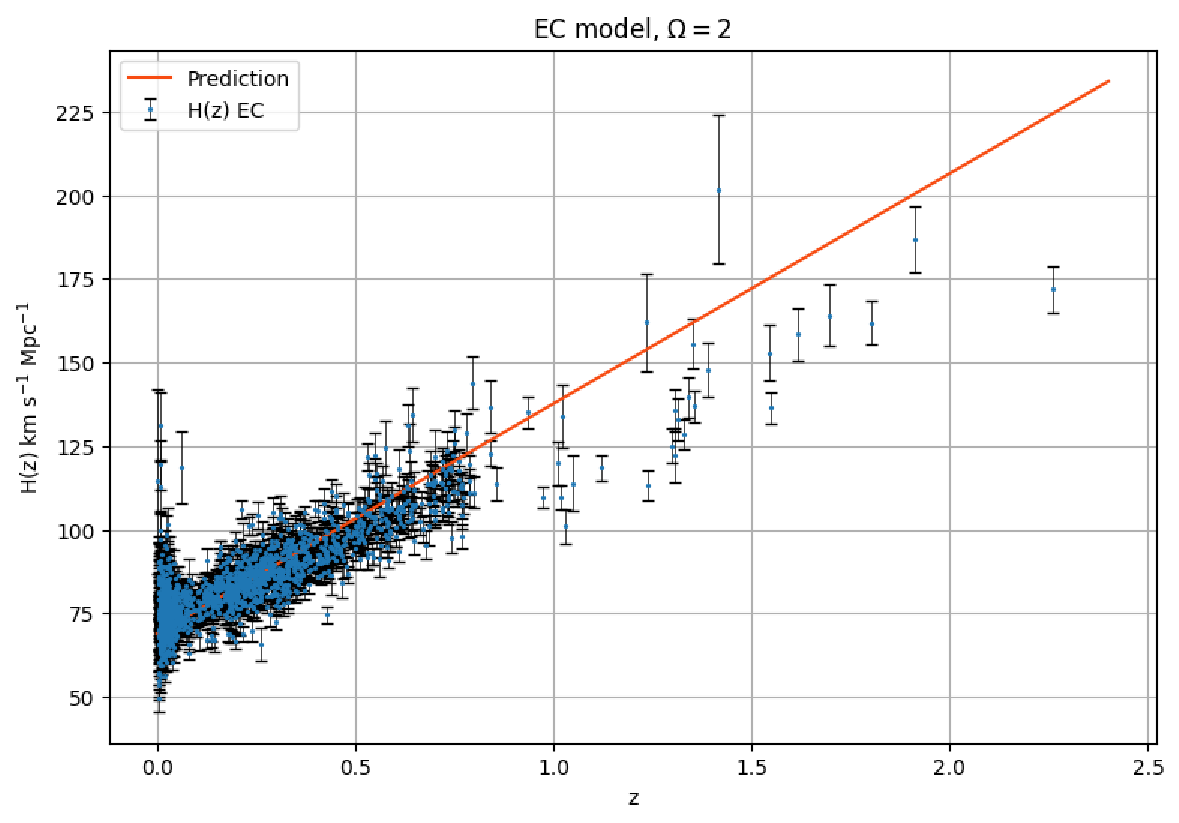}
 %  \caption{$H(z)$ in the $\Omega = 2$ case of EC model, green line gives the theoretical prediction of the model as in equation (19). The Hubble constant value estimated from MC method is $H_0 = 68.844 \pm 0.050$ km s$^{-1}$ Mpc${-1}$.}
    \label{fig:Figure5}
    \end{minipage}
 %   \hfill
    \begin{minipage}{0.45\textwidth}
        \centering
    \includegraphics[width=1\linewidth]{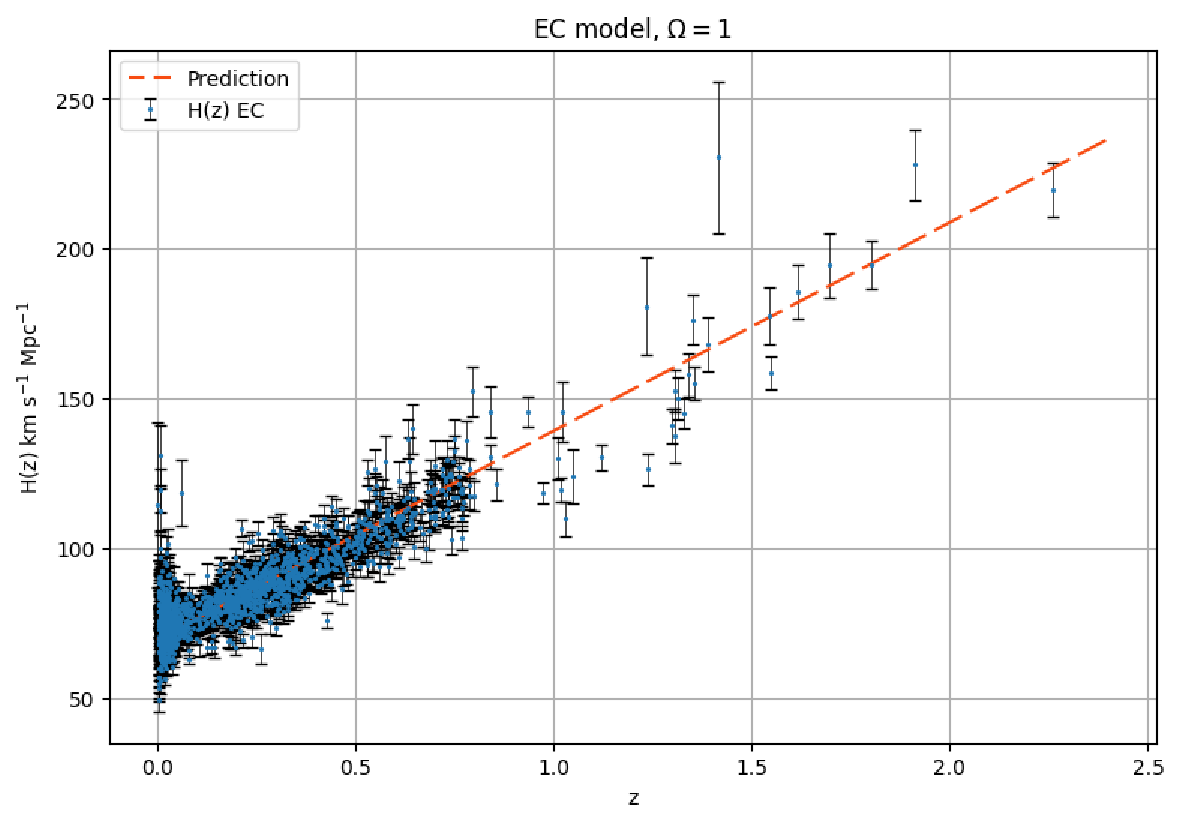}
 %   \caption{$H(z)$ in the $\Omega = 1$ case of EC model, green line gives the theoretical prediction of the model as in equation (19). The Hubble constant value estimated from MC method is $H_0 = 69.599 \pm 0.050$ km s $^{-1}$ Mpc${-1}$.}
    \label{fig:Figure4}
    \end{minipage}
%    \hfill
   \begin{minipage}{0.45\textwidth}
        \centering
    \includegraphics[width=1\linewidth]{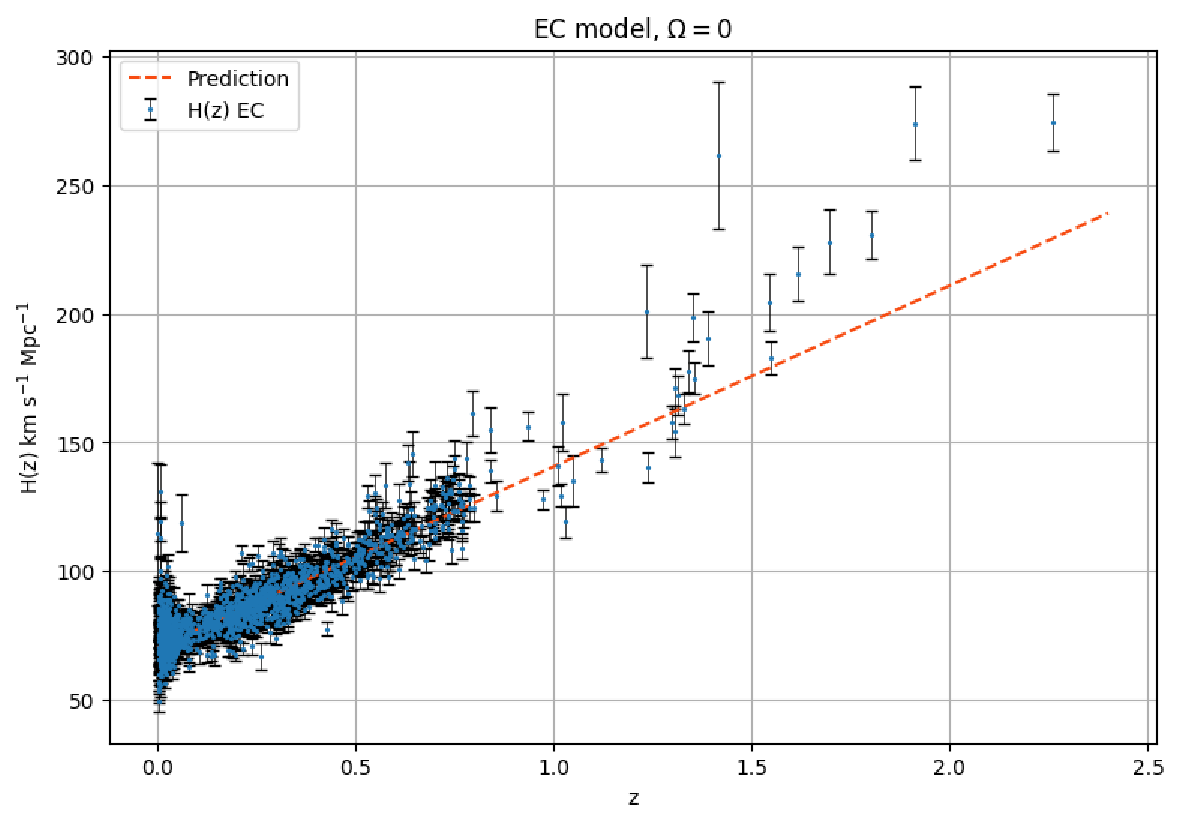}
 %   \caption{$H(z)$ in the $\Omega = 0$ case of EC model, green line gives the theoretical prediction of the model as in equation (19). The Hubble constant value estimated from MC method is $H_0 = ??? \pm 0.050$ km s$^{-1}$ Mpc${-1}$.}
    \label{fig:Figure3}
    \end{minipage}        
           \caption{$H(z)$ in the three cases of  EC model. The green lines give the theoretical prediction of the model as in equation (\ref{eq:HzEC}). The Hubble constant value estimated from regression and  MC methods is (a) $\Omega = 2$ case,  $H_0 = 68.844 \pm 0.051$ km s$^{-1}$ Mpc$^{-1}$. (b) $\Omega = 1$ case,  $H_0 = 69.599 \pm 0.050$ km s$^{-1}$ Mpc$^{-1}$.  (c) $\Omega = 0$ case, $H_0 = 70.294 \pm 0.054$ km s$^{-1}$ Mpc$^{-1}$. }
\end{figure}

 In the EC model, $\Omega$ is a free parameter, along with $H_j$ and $M$. We consider  three different values of the total density parameter; $\Omega$ = 0,  1  and 2, belonging to the three spatial geometries.  In each of these cases, we have the parameter $\alpha $ in equation (\ref{eq:mct}) equal to unity. The  set of points $\bar{H}_j$ and $\sigma_{H_j}$ for all these cases are plotted against their redshift $z_j$ in Fig. \ref{fig:Figure3}. 
  In the EC model, we have the equation for the evolution of $H(z)$ as

\begin{equation}
H(z)=H_0 (1+z) \label{eq:HzEC}.
\end{equation} 
As in the previous case of $\Lambda$CDM model, we use the regresssion and Monte Carlo methods to find out the best value of $H_0$. We have overplotted the straight line in the above equation with this value of $H_0$.

\subsection{Scatter in the $H(z)$ diagram}
After the epoch-making release of SNe Ia data in 1998, very stringent attempts were made to reduce the observational errors in the $m$-$z$ data of these objects. However, it now appears that these errors cannot be reduced any further. This  raises the question whether the scatter in the Hubble diagram is truly due to systematic or random errors, or whether it is due to any inherent property of the cosmic evolution, as  that of a fluctuating expansion  for the universe.

\begin{figure}
    \centering
    \begin{minipage}{0.45\textwidth}
       \centering
    \includegraphics[width=1\linewidth]{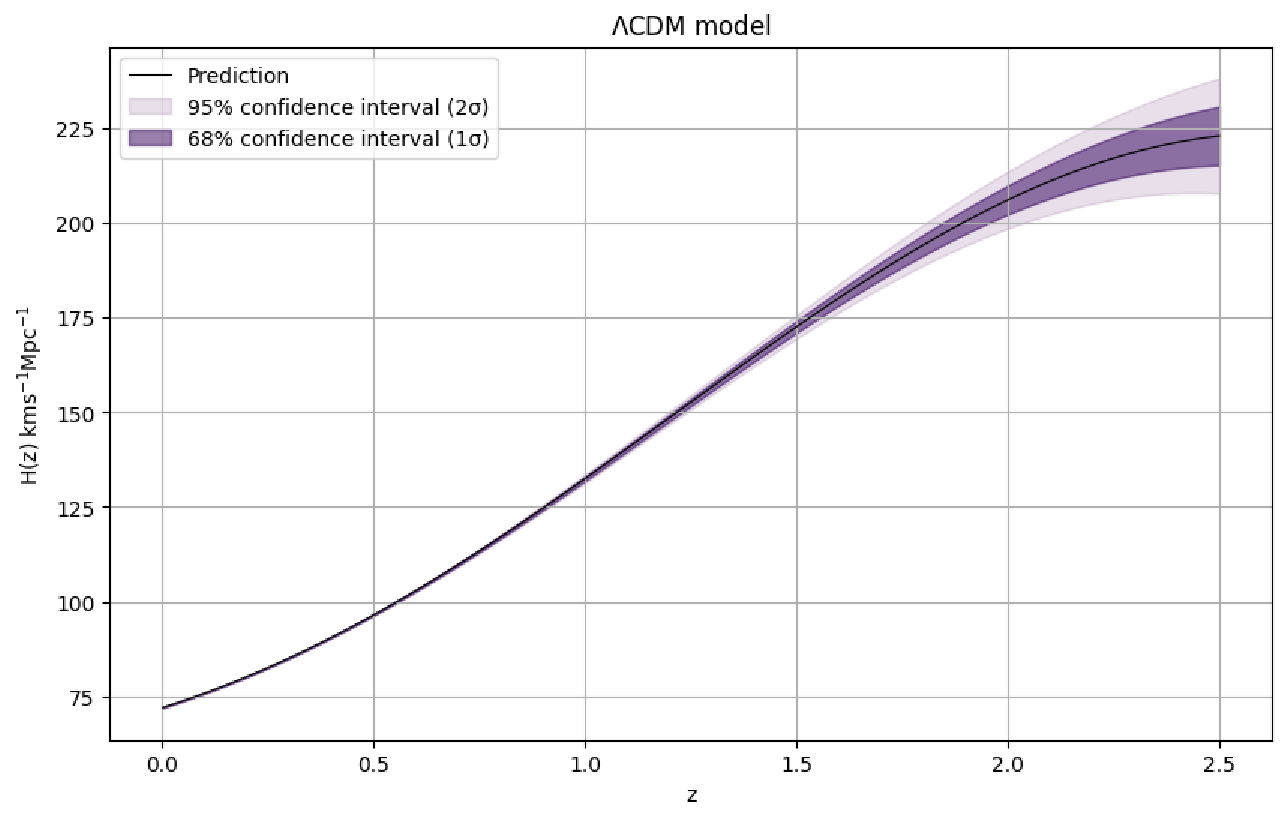}
    \caption{Evolution of scatter $\sigma_H$ in the $\Lambda$CDM model from the GPR method.}
    \label{fig:Figure6}
    \end{minipage}
    \hfill
    \begin{minipage}{0.45\textwidth}
        \centering
    \includegraphics[width=1\linewidth]{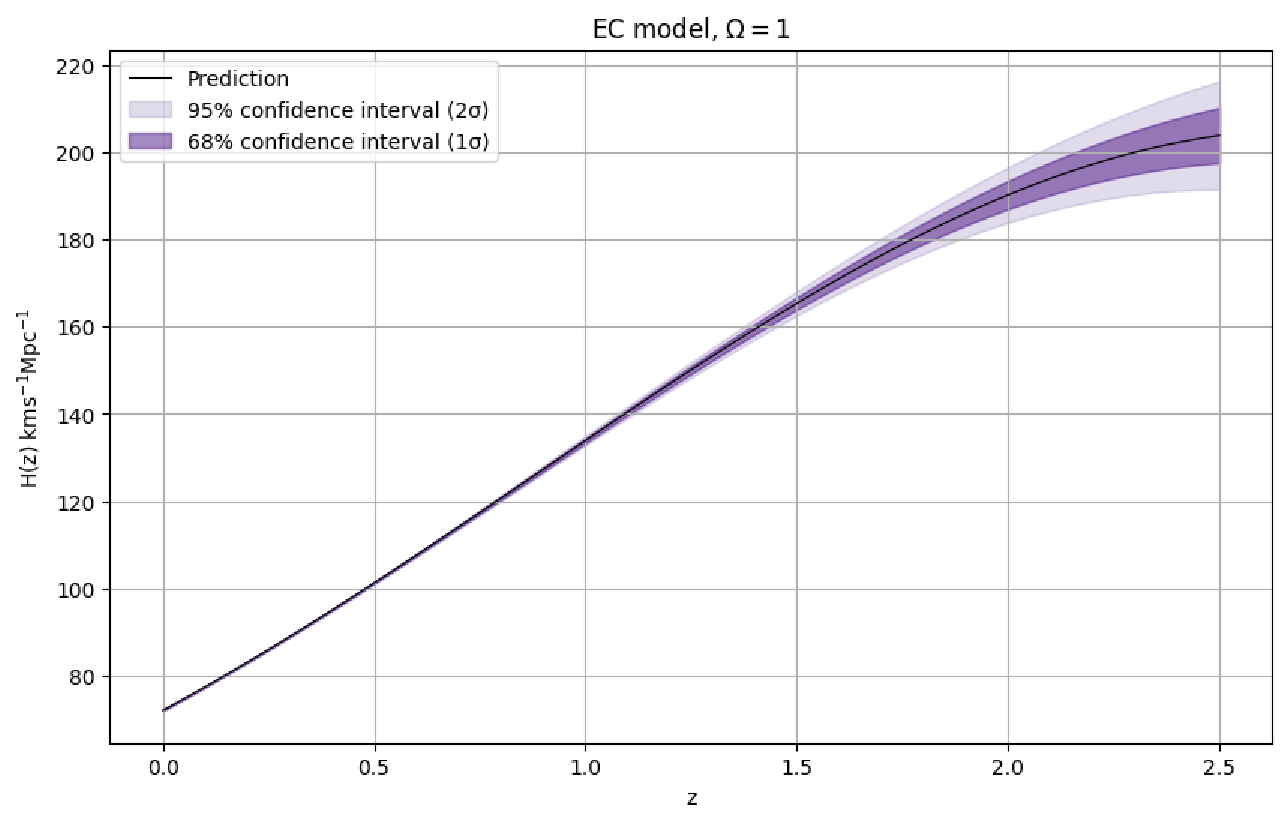}
    \caption{Evolution of scatter $\sigma_H$ in the $\Omega = 1$ case of EC model using the GPR method.}
    \label{fig:Figure7}
    \end{minipage}
\end{figure}

In this work, we have also made an attempt towards exploring the nature of the scatter in the Hubble diagram, as provided by the Pantheon+ data. The above analysis of latest Pantheon+ dataset shows that the scatter $\sigma_H$ associated with the Hubble parameter increases with increasing redshifts. This was explicitly seen in our analysis of the $H(z)$ diagram discussed above, while using Gaussian process regression (GPR). Such an evolution is exhibited by all the models considered -- the $\Lambda$CDM and the three cases of EC. The fluctuating behaviour of $H(z)$ points to some non-deterministic expansion of the universe, especially in the early epochs. This indicates that the $H$-$z$ diagram we constructed from the Pantheon+ data  can be a potential observational tool in searching for any fluctuating stochastic evolution of the early universe. Here we recall that such a possibility is discussed in \citep{berera1994stochastic,sivakumar2001stochastic,john2003classical}. The results we obtained in the present work is  depicted using the GPR method, in  Fig. \ref{fig:Figure6} and Fig. \ref{fig:Figure7}, where the former  gives the evolution of scatter in the $\Lambda$CDM model and the latter gives the same for the case of the $\Omega = 1$  EC model. The expansion rate characterised by the $H(z)$ values approach a deterministic character in the late epochs with low redshifts, but they show  considerable fluctuation and uncertainty in the early epochs. 

\section{Cosmic chronometers and comparison of cosmological models}

The differential age method  itself is a very promising tool to explore the cosmic expansion history, in a model-independent way.  In this approach, one rewrites the Friedmann equation to obtain

\begin{equation}
H(z)=-\frac{1}{1+z}\frac{\mathrm{d}z}{\mathrm{d}t}
\end{equation}
The set of data called cosmic chronometers helps to evaluate the differential age $\mathrm{d}t/\mathrm{d}z$ of the universe at different redshifts, and hence to compute the Hubble parameter $H(z)$ at various cosmic times, using this equation. In any given cosmological model, where a solution $a(t)$ is available, one has a prediction for $H(z)$ and a comparison of this with values estimated from  the cosmic chronometer data can be done to test the model.
In the $\Lambda$CDM model, the expression for $H(z)$ is as given in equation (\ref{eq:Hzlcdm}) and 
in the EC model, the corresponding equation for $H(z)$ is as given in equation (\ref{eq:HzEC}). In this section, we make an attempt for a Bayesian model comparison of the four models considered here, among themselves, using the cosmic chronometer data in \citep{moresco2024addressing}.

\subsection{Bayesian model comparison} \label{subsec:bayesian}

The use of Bayes theorem, without a proper account of the use of prior probability in it, may lead to erroneous conclusions. Bayes theorem helps to evaluate the posterior probability $P(M_1|D, I)$ for a model $M_1$ (that is, the probability for a model $M_1$ to be true, given the data $D$ and the truth of some background information $I$) and   can be stated as

\begin{equation}
P(M_1|D, I)=\frac{P(M_1|I) P(D|M_1, I)}{P(D|I)}
\end{equation}
Here $P(M_1|I)$ is the prior probability, which we assume before considering the data, $P(D|M_1, I)$ is the probability to get the data $D$ if the model $M_1$ and the background information $I$ are true, and $P(D|I)$ is a normalisation constant, which is the probability for the data $D$, whatever be the model. The prior probability is an educated guess of the probability for the model, given only the background information $I$. $P(D|M_1, I)$ is sometimes called the likelihood for the model $M_1$ and is denoted as ${\cal L}(M_1)$. One can see that $P(D|I)$, the normalisation constant, is the sum of posteriors of all possible models such as $M_1$. When $M_i$ are models or hypotheses, it is often impossible to evaluate it, but  this will not be a major impediment in Bayesian model comparison.   When we compare two models $M_1$ and $M_2$, this quantity cancels out on taking the ratio and one gets the Bayesian odds 

\begin{equation}
O_{12}\equiv \frac{P(M_1|D, I)}{P(M_2|D, I)}=\frac{P(M_1|I) P(D|M_1, I)}{P(M_2|I) P(D|M_2, I)}
\end{equation}
If the background information $I$ does not give any preference to one model over the other, the two priors are equal and the above ratio simply becomes the Bayes factor $B_{12}$

\begin{equation}
B_{12}\equiv \frac{ P(D|M_1, I)}{ P(D|M_2, I)} = \frac{{\cal L}(M_1)}{{\cal L}(M_2)},
\end{equation}
 which is the ratio of the two likelihoods. The problem of the estimation of the likelihood $P(D|M_1, I)\equiv {\cal L}(M_1)$  is solved by  using the Bayes theorem once again. The  posterior probability for a parameter in  $M_1$ to have a value $\theta$, given the data $D$ and also given the truth of both the model $M_1$ and the background information $I$, can be written using Bayes theorem as

\begin{equation}
 P(\theta|D,M_1,I)= \frac{P(\theta |M_1,I)P(D|\theta, M_1, I)}{P(D|M_1, I)}
\end{equation}
Here, the denominator, which is a normalisation constant, can be evaluated as an integral of $ P(\theta | D, M_1, I)$  over all the possible values of $\theta $  in this model. This is the probability for the data $D$, given the model $M_1$ and the information $I$, which is the desired likelihood ${\cal L}(M_1)$ for the model. Thus

\begin{eqnarray}
{\cal L}(M_1) &\equiv & P(D|M_1, I) \nonumber \\
&=&\int P(\theta | D, M_1, I) \mathrm{d}\theta \nonumber \\
&= &\int P(\theta |M_1,I)P(D|\theta, M_1, I) \mathrm{d}\theta \label{eq:likeli_theta} 
\end{eqnarray}
$P(D|\theta, M_1, I)$, the likelihood for the parameter $\theta$, given $D$, $M_1$ and $I$ are true, is usually in the form of the $\chi^2$ statistic, as in equation (\ref{eq:chi2prob}).

The first factor in the integrand on the right hand side of  equation (\ref{eq:likeli_theta}) is the prior probability for the parameter $\theta$ in model $M_1$. Before we start to analyse the data, i.e., when we have only the background information $I$,  there may exist some consensus on the range of values of the parameter $\theta$, or even a PDF for $\theta$, in   the model.  Often this  is the result of a previous analysis. While estimating the Bayesian probability, the prior is intended as the posterior obtained from a previous analysis of data, in a specific model. When this can be approximated as a Gaussian distribution with mean $\mu$ and standard deviation $\sigma$, one can write the prior probability for the parameter $\theta$  in the form 

\begin{equation}
 P(\theta|M_1,I)= \frac{1}{\sqrt{2\pi \sigma^2}}\exp \left(-\frac{(\theta-\mu)^2}{2\sigma^2}\right)
\end{equation}

The above discussion points to the importance of including the background information $I$ in Bayes theorem, while evaluating the posterior probability for a model.  The background information, which is commonly shared by all before the analysis of the data, has an equal contribution in deciding the prior probability as that information one has deduced directly from the model in a previous analysis. This effectively  means that the prior probability distribution one assigns for a parameter in a model should lie within the commonly accepted range of values  of that parameter. This discussion also  underscores the fact that it is not appropriate to decide on prior probabilities  on the basis of the same data under analysis.

\subsection{Model comparison with the full set of CC data}

We shall now perform  Bayesian model comparison of the $\Lambda$CDM model with each of the three coasting models (with  $\Omega$ = 0, 1 and  2) and also among the three EC models, using the CC data. The latest compilation of CC data, which we shall use here is from \citep{moresco2024addressing}.

\begin{figure}
    \centering
    \includegraphics[width=1\linewidth]{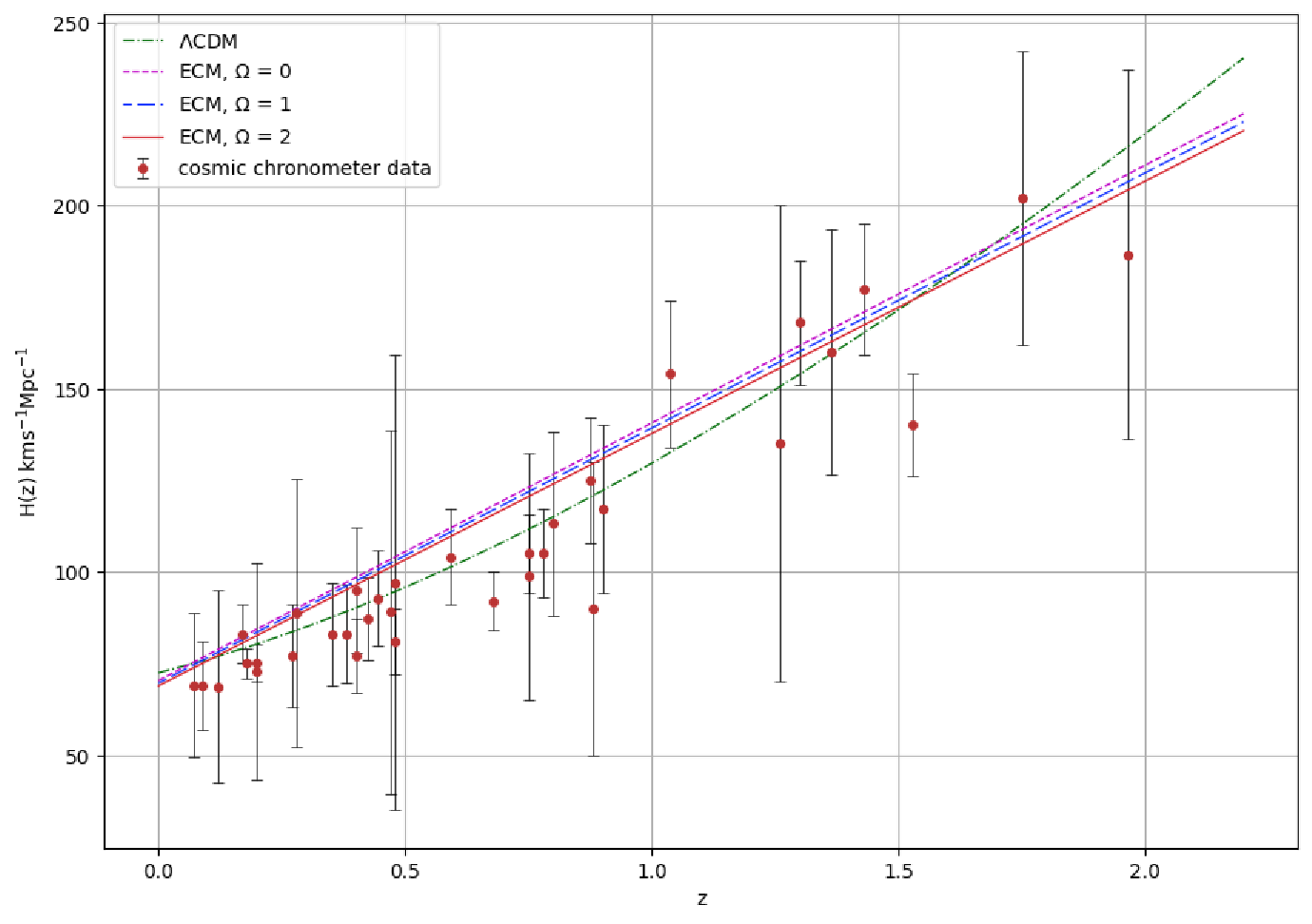}
    \caption{The full set of 35 cosmic chronometer data points with errorbars. The best-fitting $H(z)$ curves  corresponding to the $\Lambda$CDM model and the three EC models are plotted.}
    \label{fig:35CCpoints}
\end{figure}

Now onwards, model $M_1$ refers to $\Lambda$CDM model, $M_2$ to the $\Omega = 2$ case, $M_3$ to the $\Omega = 1$ case and $M_4$ to the $\Omega = 0$ case of EC model. The likelihood for each model is evaluated after setting the prior probabilities for the parameters in them as discussed in the above subsection. The prior probabilities for $H_0$ shall be Gaussian functions in the respective models, with mean and standard deviation as evaluated from the Pantheon+ data, discussed in the previous section. In the case of parameter $\Omega_m$ appearing in the $\Lambda$CDM model, we adopted a Gaussian function with  \(\Omega_m = 0.315  \pm  0.007\)  \citep{lahav2024cosmological} as prior. Bayes factors between different models, which we have evaluated, are given in Table \ref{tab:B35CC}.

The large value of Bayes factor $B_{12}$ between $\Lambda$CDM model and the $\Omega =2$ case of EC model  indicates very strong evidence in favour of the former over the latter. Bayes factors $B_{23}$, $B_{24}$, $B_{34}$ are the results of Bayesian comparison among EC models themselves. Comparatively larger values of these factors can be interpreted as  favouring the $\Omega = 2$ model over the $\Omega = 1$ ($R_h = ct$) and $\Omega = 0$ [Milne-type \citep{milne1935,milne1948}] models.

\begin{table} 
\centering
\begin{tabular}{c c}
    \hline
    &  Bayes factor \\
    \hline
     $B_{12}$ &  10466.98 \\
     $B_{24}$ & 264.70 \\
     $B_{23}$ & 15.96  \\
     $B_{34}$ &  16.59  \\
    \hline 
\end{tabular}
\caption{\label{tab:B35CC} Bayes factors estimated from the comparative study between models using 35 CC data points. }
\end{table}

The method adopted in \citep{melia2013cosmic,melia2018model} of using the prior
around \(63.2 \pm 1.6 \) km s$^{-1}$ Mpc$^{-1}$ for the $R_h=ct$ model is  unacceptable, since no other  observations in the present universe help us to motivate a prior of this kind. A straight line as in equation (\ref{eq:HzEC}) is fitted to the CC data  and they obtained a value of $H_0$ from the best-fitting line as $H_0 = 63.2\pm 1.6$ km s$^{-1}$ Mpc$^{-1}$ which is in sharp tension with other estimations of $H_0$, such as that obtained from the SNe Ia data, in the same model.  Moreover, a prior cannot be chosen from the same data we want to analyse, as done in their analysis.  In view of the discussion in Subsec. \ref{subsec:bayesian}, we can conclude that the analysis in \citep{melia2013cosmic,melia2018model} is not a truly Bayesian one. Additionally, fitting a straight line as in the above case will give only one value of $H_0$ for the EC model, and it would then be irrespective of the value of $\Omega$ in the model. In other words, it is not clear to which value of $\Omega$ such  $H_0$ belongs.

\subsection{Model comparison with a reduced dataset}
Any dataset may contain outliers. It is described by National Institute of Standards and Technology (.gov) thus:

`An outlier is an observation that lies (at) an abnormal distance from other values in a random sample from a population. In a sense, this definition leaves it up to the analyst (or a consesus process) to decide what will be considered abnormal.' (https://www.itl.nist.gov/div898/handbook/prc/

section1/prc16.htm) 

Let us choose any  point in the  dataset which lie far away from the theoretical curves under study  as an outlier. Outliers in data are usually culpable in the context of statistical tests, regression and parameter estimation. In this subsection, we investigate the result of excluding some such outlier points from the CC data, based on a standard criterion and see whether that affects our above conclusions drastically.

\begin{table}
\centering
\begin{tabular}{c c c}
    \hline
    $z$ & $H(z)$ & References \\
    \hline
    0.179 & $75 \pm 4$ & \citep{moresco2012improved} \\
    0.199 & $75 \pm 5$ & \citep{moresco2012improved} \\
    0.4783 & $80.9 \pm 9$ & \citep{moresco20166} \\
    1.53 & $140 \pm 14$ & \citep{simon2005constraints} \\
    \hline 
\end{tabular}
\caption{\label{tab:CC_eliminated} The four outlier points excluded from the full CC dataset. $H(z)$ in the unit of km s$^{-1}$ Mpc$^{-1}$.}

\end{table}

Here we  treat those CC data points, which lie at more than \(1\sigma\) away from any of the theoretical curves compared in this study, as  outliers. Accepting this criterion, we  identify four data points from \citep{moresco2012improved, moresco20166, simon2005constraints}. These points, given in Table \ref{tab:CC_eliminated}, are eliminated from the original CC data  of 35 data points in \citep{moresco2024addressing}.   The Bayesian model comparison using the remaining 31 CC data points gives Bayes factors, which are   given in Table \ref{tab:B31CC}. The results show drastic decrease in the Bayes factors than that in Table \ref{tab:B35CC}. 

We argue that this considerable reduction in the Bayes factor values $B_{12}$, $B_{24}$, $B_{23}$, $B_{34}$  indicates an  incompatibility existing between the SNe Ia data and CC data. This is particularly for the reason that none of the other Bayesian model comparisons performed so far \citep{john2002comparison,john2010bayesian,melia2018model} between the $\Lambda$CDM and EC models using SNe Ia have given  values of Bayes factor as large as that in  Table \ref{tab:B35CC}.

\begin{table} 
\centering
\begin{tabular}{c c}
    \hline
    & Bayes factor \\
    \hline
     $B_{12}$ & 50.20   \\
     $B_{24}$ &  19.78  \\
     $B_{23}$ &  4.26 \\
     $B_{34}$ &  4.64 \\
    \hline 
\end{tabular}
\caption{\label{tab:B31CC} Bayes factors from the comparative study of models using 31 CC data points.}
\end{table}

\section{Conclusion}
The $\Lambda$CDM model, albeit being a successful one on several fronts,  faces challenges such as  the Hubble tension \citep{di2022challenges}. The studies we perform in this work are particularly important in the context of the Hubble tension.

Well-known model-independent  estimations of the cosmic evolution are  based on observations of  the CC and  gravitational waves (GW) \citep{raffai2024constraints}.  The GW waves emitted by merging black holes and neutron stars, which are termed standard sirens, were recently used to  constrain cosmological parameters and to compare  different cosmological models, including the EC models. In the first test of coasting models using  GW sirens \citep{raffai2024constraints}, constraints were put on $H_0$, the present value of the Hubble parameter, for three fixed values of the total density parameter $\Omega_0 =$ 0, 1 and 2 (corresponding to  $k=$ -1, 0 and +1, respectively). In a Bayesian model comparison,  it was found that the coasting models and the $\Lambda$CDM models fit equally well to the applied set of GW detections.   Moreover, they have found that the maximum posterior probability for the $k=+1$ EC model is the closest to the $H_0$ measured by the differential age method of the CC data.

The evolution of the Hubble parameter that we estimated in this work, for the $\Lambda$CDM model and the three special cases of EC model using the SNe Ia data, offers a new perspective to look into the expansion history of universe. The growth of scatter as we go to the early epochs in both models is strong evidence for suspecting some stochastic nature of the Hubble parameter. The present value of the Hubble parameter inferred from our study, in units of km s$^{-1}$ Mpc$^{-1}$, are $H_0 = 72.391 \pm 0.053$ for the $\Lambda$CDM model and $H_0=68.844 \pm 0.051$, $H_0=69.599 \pm 0.050$ and $H_0=70.294 \pm 0.054$, for the $\Omega=2$, $\Omega=1$ and $\Omega=0$ cases of EC model, respectively. The fact that CC data is  model-independent makes it ideal for comparative study between cosmological models.  The results of Bayesian model comparison strongly supports the $\Lambda$CDM model over other EC models, while using the full dataset of 35 CC.  Among the EC models, $\Omega = 2$ case, which corresponds to closed spatial geometry (\(k  = +1\)), is favoured over $\Omega = 1$ (\(R_h = ct\)) and $\Omega = 0$ (Milne-type) models, in both of our comparative studies.  However, large reduction happened in the Bayes factor between all these models from the exclusion of mere four outlier points. In the context of having only marginal support for $\Lambda$CDM model over the EC models in several previous analyses of SNe and GW data, we conclude that the large values of Bayes factor obtained while using the full set of 35 CC data points  indicate an incompatibility between SNe Ia data and the CC data.

\section*{Acknowledgements}
We are grateful to Juan Casado for a critical reading of the manuscript and for  valuable comments.

\section*{Data Availability}
The Pantheon+ dataset used in this study can be found at \href{https://github.com/PantheonPlusSH0ES/DataRelease}{https://github.com/PantheonPlusSH0ES/DataRelease} and the cosmic chronometer data are from \citep{moresco2024addressing}.

\label{lastpage}

\bibliographystyle{mnras}
\bibliography{cite}

\end{document}